\documentclass{PoS}

\title{Identification of newly-discovered sources belonging to the 4th IBIS catalog and to the 54 months Palermo Swift/BAT catalog}

\ShortTitle{New AGN classifications}

\author{{Pietro Parisi$^{1,2}$, N. Masetti$^2$, A.F. Rojas$^3$, V. McBride$^{4,5}$, L. Steward$^{4,5}$, L. Bassani$^2$, A. Bazzano$^1$,  A.J. Bird$^6$, P.A. Charles$^{4,6}$, V. Chavushyan$^7$, G. Galaz$^3$, E. Jim\'enez-Bail\'on$^8$, R. Landi$^2$, A. Malizia$^2$, E. Mason$^9$, D. Minniti$^{3,10}$, L. Morelli$^{11}$, E. Palazzi$^2$, J.B Stephen$^2$ and P. Ubertini$^1$}\\
       $^1$INAF -- IAPS Roma, Rome, Italy\\
       $^2$INAF -- IASF Bologna, Bologna, Italy\\
       $^3$Pontificia Universidad Cat\'olica de Chile, Santiago, Chile\\
       $^4$South African Astronomical Observatory, Cape Town, South Africa\\
       $^5$ACGC, University of Cape Town, Rondebosch, South Africa\\
       $^6$University of Southampton, UK\\
       $^7$Instituto Nacional de Astrofis\'ica, \'Optica y Electr\`onica, Puebla, M\'exico\\
       $^8$Universidad Aut\'onoma de M\'exico, M\'exico DF\\
       $^9$Space Telescope Science Institute, Baltimore, USA\\
       $^{10}$Specola Vaticana, Citt\`a del Vaticano\\
       $^{11}$Dipartimento di Astronomia, Universit\`a di Padova, Italy\\
        E-mail: \email{pietro.parisi@iaps.inaf.it}}

\abstract{The most recent all-sky surveys performed with the INTEGRAL and SWIFT satellites allowed the detection of more than 1500 sources in hard X-rays above 20 keV. About one quarter of them has no obvious counterpart at other wavelengths
and therefore could not be associated with any known class of high-energy emitting objects. Although cross-correlation with catalogues or surveys at other wavelengths (especially soft X-rays) is of invaluable support in pinpointing the putative optical candidates, only accurate optical spectroscopy can reveal the true nature of these sources.
With the aim of identifying them, we started in 2004 an optical spectroscopy program which uses data from several telescopes worldwide and which proved extremely
successful, leading to the identification of about 200 INTEGRAL objects and nearly 130 Swift sources.
Here we want to present a summary of this identification work and an
outlook of our preliminary results on identification of newly-discovered sources belonging to the 4th IBIS catalog and to the 54 months Palermo Swift/BAT catalog.}

\FullConference{An INTEGRAL view of the high-energy sky (the first 10 years)"
9th INTEGRAL Workshop and celebration of the 10th anniversary of the launch,\\
		October 15-19, 2012\\
		Bibliotheque Nationale de France, Paris, France}

\begin{document}

\section{Introduction}
Thanks to the {\it INTEGRAL} and {\it Swift} satellites the way of looking at the hard X-ray sky above 20 keV has changed substantially. Through the unique imaging and spectroscopy capabilities of the IBIS and BAT instruments that are the basis of the {\it INTEGRAL} and {\it Swift} surveys ([1], [2], [3], [4]), these satellites have improved the knowledge on hard X-ray sources in terms of sensitivity and positional accuracy. Many of the sources belonging to these surveys are however of unidentified nature, but the combined use of available information at longer wavelengths (mainly soft X-rays and radio) and, above all, optical spectroscopy on their putative counterparts can reveal their exact nature of these hard X-ray objects. Since 2004 our group identified about 200 {\it INTEGRAL} sources and 130 {\it Swift} sources, drastically reducing the percentage of unidentified objects in the various IBIS and BAT surveys and allowing statistical studies on them. Here we present a summary of this follow-up work and an outlook of our results on the identification of newly-discovered sources belonging to the 4th IBIS and the 54 month Palermo {\it Swift}/BAT catalogs.

\section{Optical analysis}
The optical spectroscopic data presented here were obtained with various telescopes located in the Southern and Northern hemispheres.

The optical counterpart positions are pinpointed through correlations with soft X-ray and/or radio catalogues (Chandra, XMM-Newton, ROSAT, Swift/XRT, NVSS, SUMSS, MGPS-2) or specific X-ray pointings [12],  in order to reduce the position uncertainties from about 4 arcmin to less than 10 arcsec. Then we superimposed on the DSS-II-Red survey\footnote{\tt http://archive.eso.org/dss/dss/} the soft X-ray and/or radio error box and selected all the likely optical counterparts inside it (usually less than 3; [13]-[23]). Finally, we spectroscopically observed all of them.

The data reduction of their optical spectra was performed with the standard procedure (optimal extraction; [7]) using IRAF\footnote{
IRAF is the Image Reduction and Analysis 
Facility made available to the astronomical community by the National 
Optical Astronomy Observatories, which are operated by AURA, Inc., under 
contract with the U.S. National Science Foundation. It is available at 
{\tt http://iraf.noao.edu/}}.
Calibration frames (flat fields and bias) were taken on the day preceding or following 
the observing night. The wavelength calibration was obtained using lamp spectra 
acquired soon after each on-target spectroscopic acquisition. The uncertainty on the 
calibration was $\sim$0.5~\AA~for all cases. This was checked using the positions of 
background night sky lines. Flux calibration was performed using 
catalogued spectrophotometric standards.
Objects with more than one observation had their spectra stacked 
together to increase the signal-to-noise ratio.

The identification and classification approach adopted in the analysis of the optical spectra is the same applied in the papers of [13]-[26].
\section{{\it INTEGRAL}/IBIS classifications}
\begin{figure*}[th!]
\begin{center}
\includegraphics[width=11.5cm,angle=0]{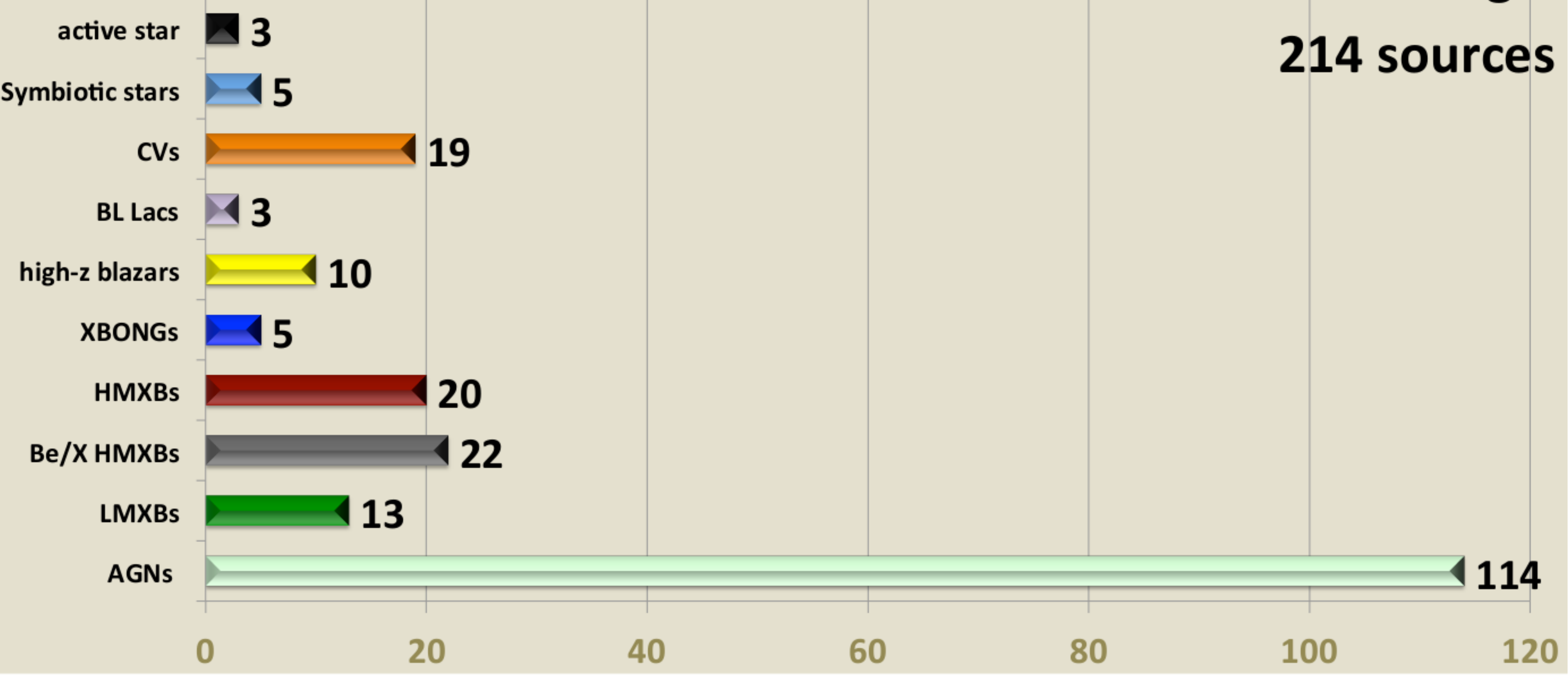}
\includegraphics[width=11.5cm,angle=0]{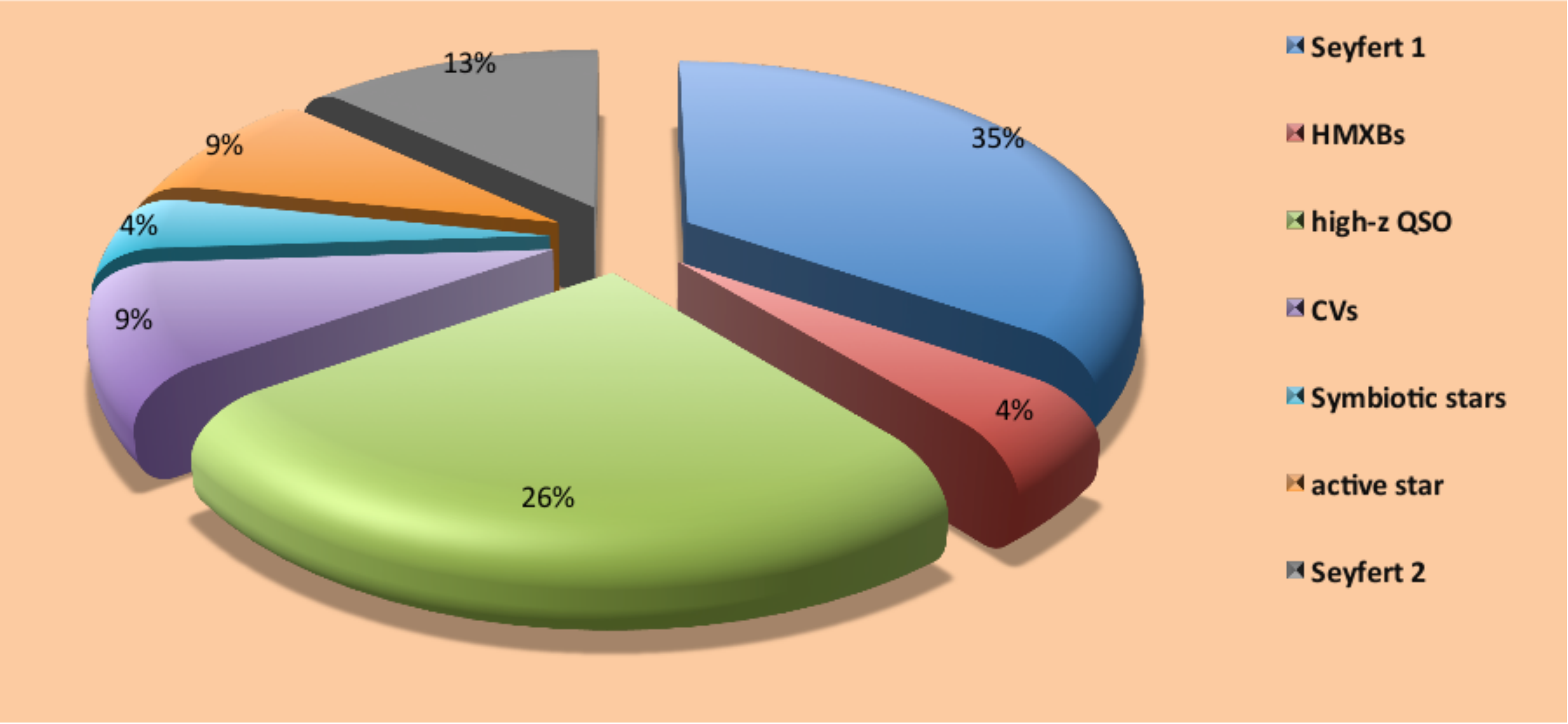}
\caption{{\it Top panel:} Histogram showing all the INTEGRAL objects, either without optical identification, or not well explored or without published optical spectra, belonging to the IBIS surveys and studied up to now by our group. {\it Bottom panel:} the 22 new identifications mainly belonging to the 4th IBIS catalog}\label{int}
\end{center}
\end{figure*}

Up to now within our {\it INTEGRAL} sources identification program we produced 11 papers ([13]-[23]). This work allowed us to identify or confirm the nature of about 200 sources. We found that the majority are Active Galactic Nuclei (AGNs, see Fig.\ref{int} for details) followed by Galactic sources.
This work allowed the detection of a large number of new AGNs, many of them in the so-called `Zone of Avoidance', i.e. along the Galactic Plane, where the presence of Galactic dust and neutral hydrogen severely hampered in the past their detection and studies at both optical and soft X-ray wavelengths. It also gave us the possibility of detecting a substantial number of new, possibly magnetic, Cataclysmic Variables (CVs; e.g. [11]; [28]).
Moreover, the future identification of sources in the bulge and southern plane would be aided through IR variability with the VVV data [24]. 

\begin{figure*}[th!]
\begin{center}
\includegraphics[width=5.5cm,angle=0]{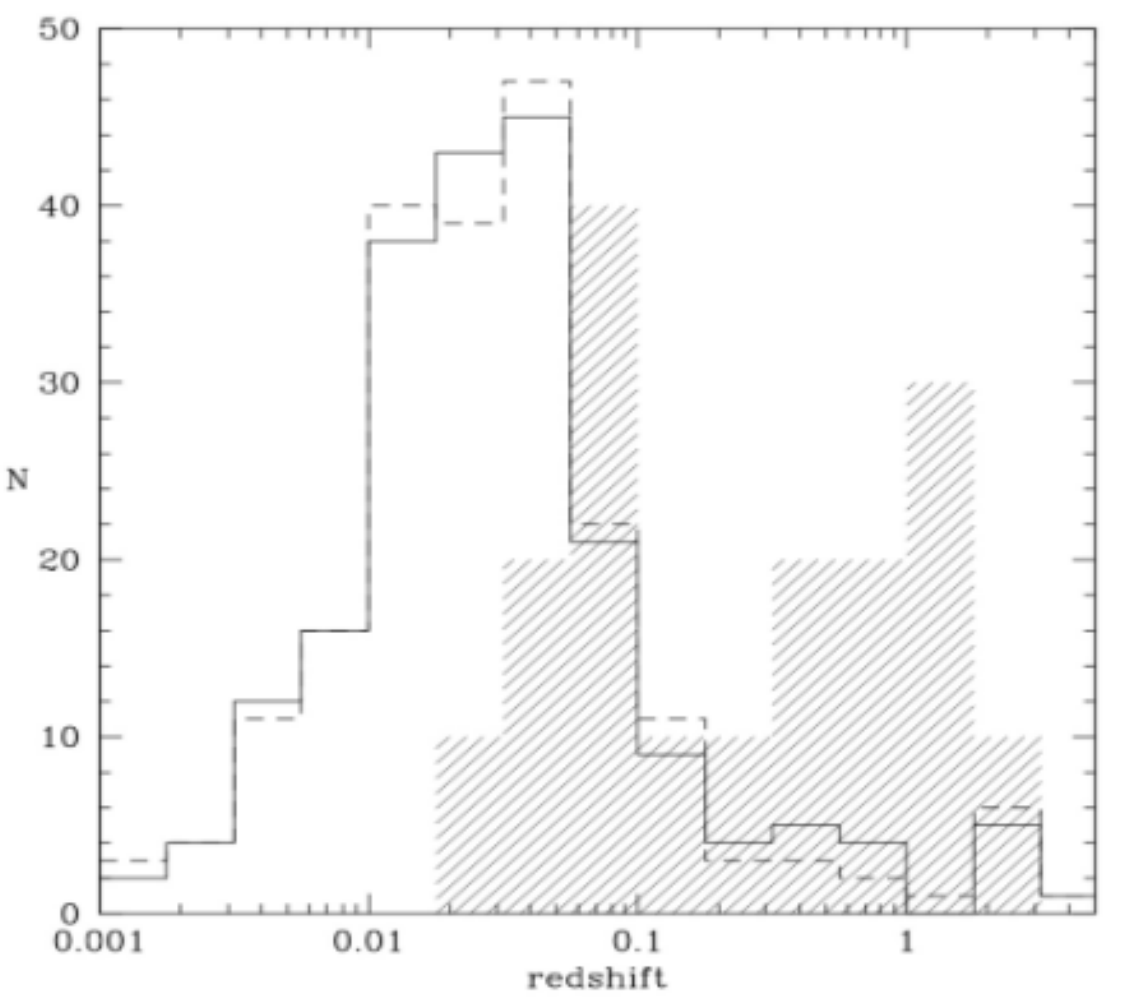}
\includegraphics[width=6.5cm,angle=0]{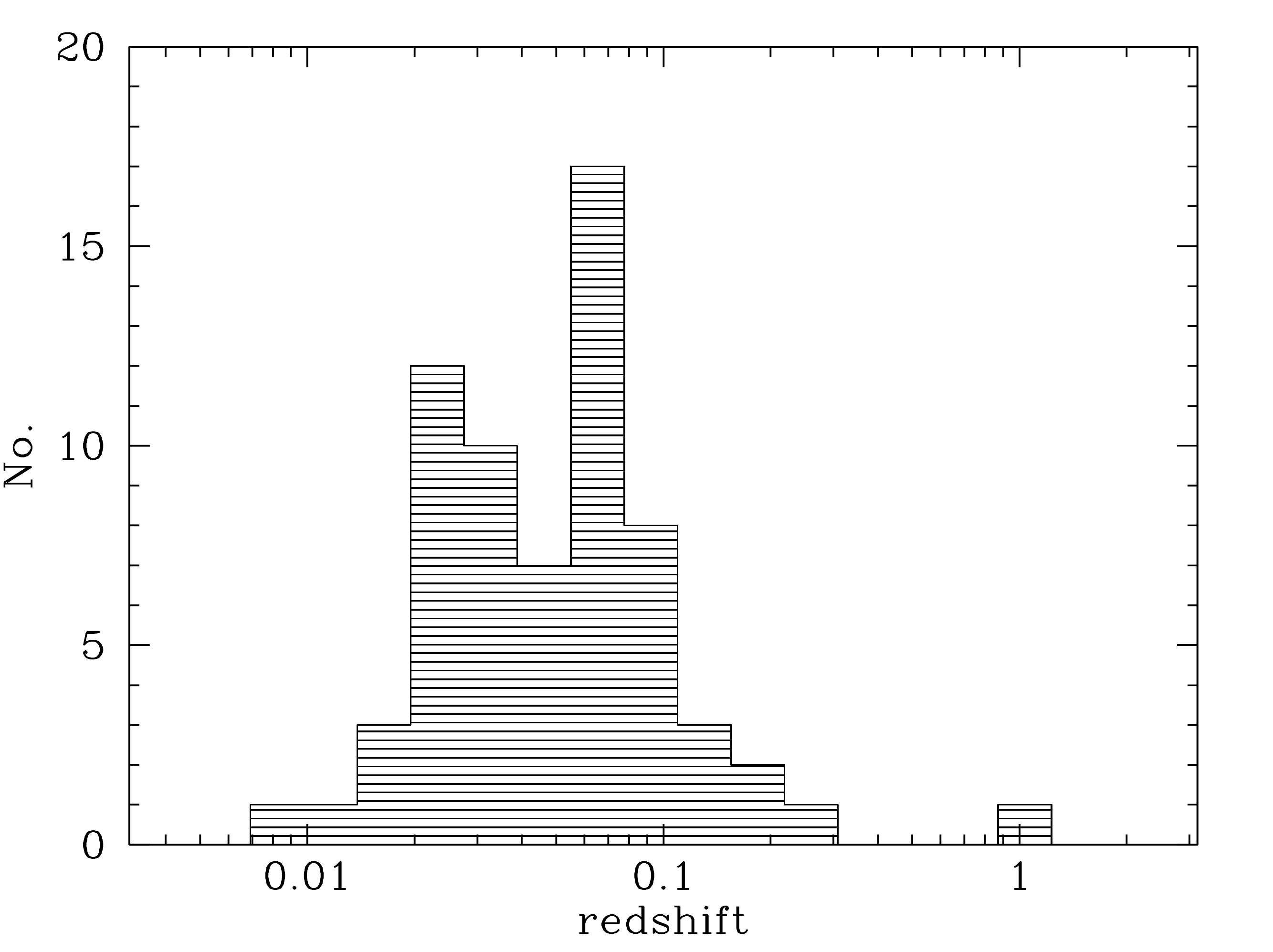}
\caption{{\it Left panel:}  Logarithmic histogram comparing the distribution of redshifts of our 22 new INTEGRAL extragalactic source identifications (shaded histogram, numbers multiplied by 10 for comparison, [23]) with those belonging to the survey of [2] (continuous lines) and [9] (dashed lines). {\it Right panel:} the redshift distributions of the 76 new identifications belonging to the 54 month Palermo {\it Swift}/BAT catalog [4].}\label{histo}
\end{center}
\end{figure*}

In the last work on the identification of 22 new {\it INTEGRAL}/IBIS sources [23] we found 8 Seyfert 1, 3 Seyfert 2, 6 high-z QSO, 1 High Mass X-ray Binary (HMXB), 2 CVs, 1 Symbiotic star and 1 active star.
For these new {\it INTEGRAL} identifications we also used medium-sized and large telescopes (4-metres class or larger), in order to study the faint end of the distribution of the putative optical counterparts of these high-energy sources. In Fig. \ref{histo}, left panel, we report the redshift distribution of our sample compared to that of [2] and [9]. As evident from the histograms the AGN redshift distribution of [23] lies at higher redshift values with respect to the AGN redshift distribution of the catalogues of [2] and [9]. Indeed the average redshift of the extragalactic objects of [23] is  $<z> = 0.576$, while those of [2] and [9] are $ <z> = 0.135$ and $<z> = 0.144$ respectively. We applied a Kolmogorov-Smirnov nonparametric test to verify our result, and we found that the probability that the AGNs of [23] and those of the two catalogs ([2] and [9]) come from the same redshift distribution is less than 0.001. This confirms that the AGNs of the sample of [23] are drawn from a different distribution, containing more distant objects compared to the other two in [2] and [9] likely thanks to the use of medium-sized and large telescopes.

\section{{\it Swift}/XRT classifications}
\begin{figure*}[th!]
\begin{center}
\includegraphics[width=11cm,angle=0]{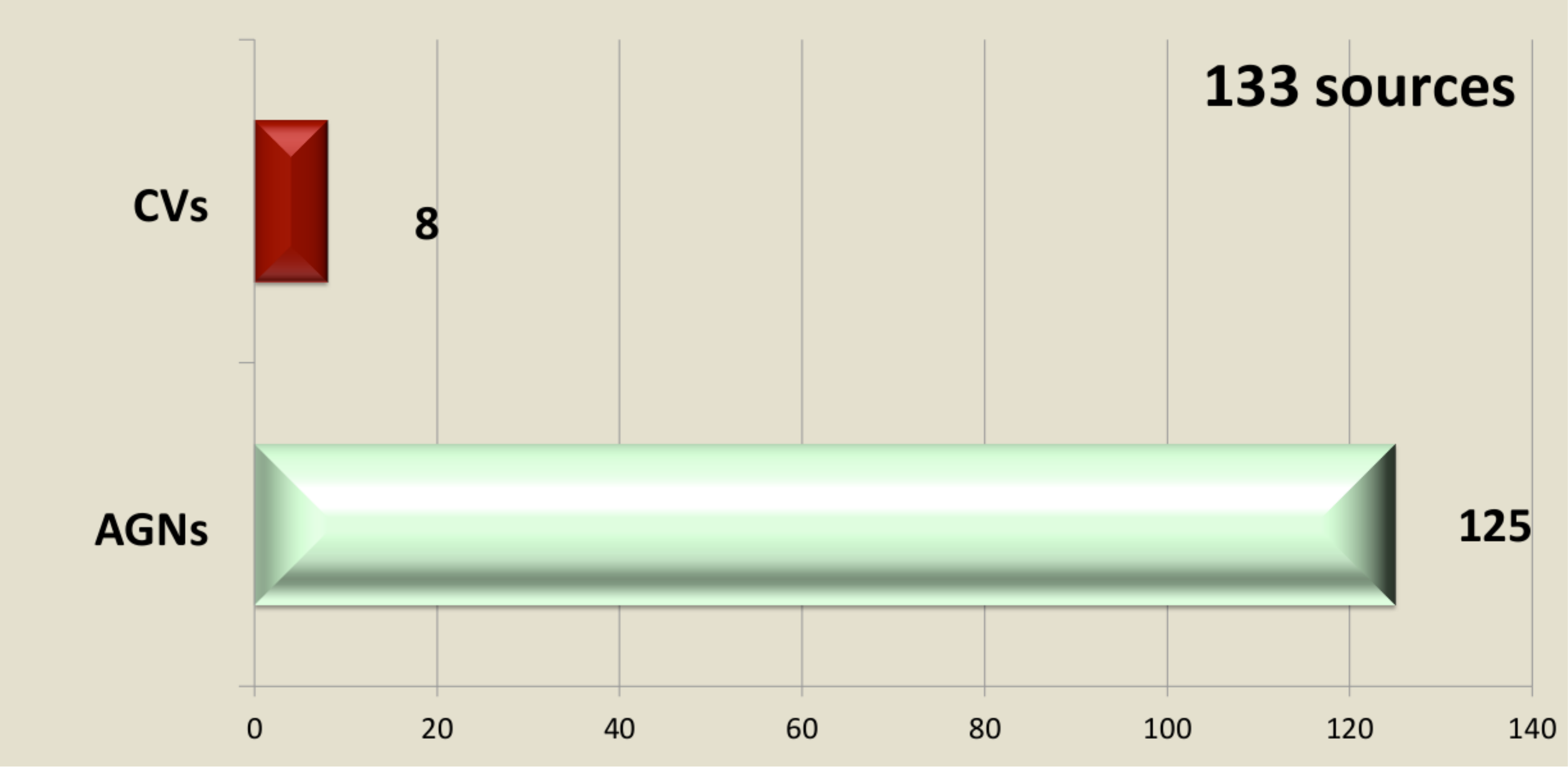}
\includegraphics[width=11cm,angle=0]{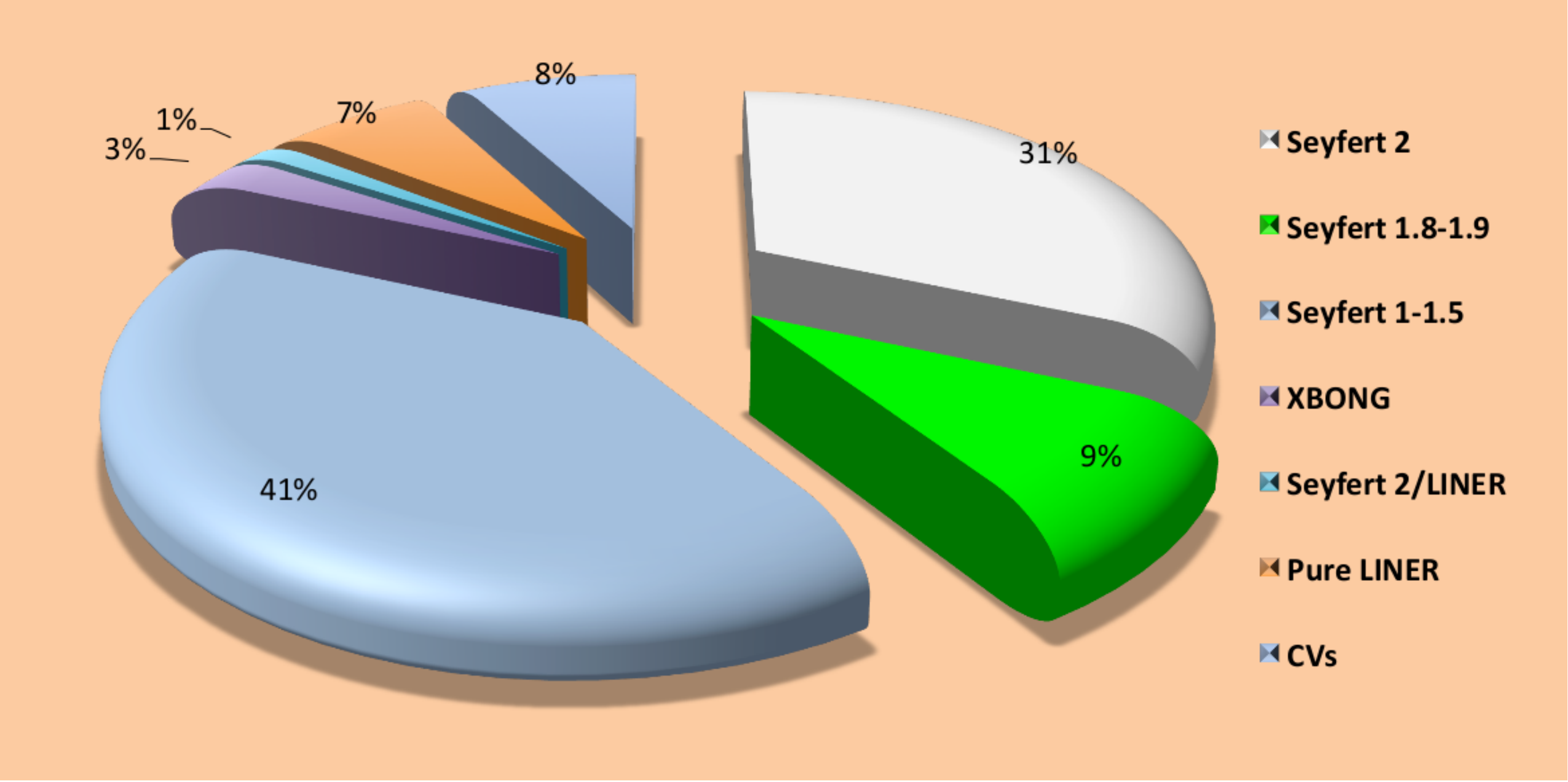}
\caption{{\it Top panel:} Histogram showing all the {\it Swift} objects, either without optical identification, or not well explored or without published optical spectra, belonging to the BAT surveys and studied up to now by our group. {\it Bottom panel:} pie chart of the 76 new identifications belonging to the 54 month Palermo {\it Swift}/BAT catalog.}\label{swi}
\end{center}
\end{figure*}

Within our {\it Swift} sources identification program we produced 3 papers ([9], [25] and [26]) and the 4th one is in preparation. We identified 133 {\it Swift}/BAT sources, 125 AGNs and 8 CVs (see Fig. \ref{swi} for details), belonging to the BAT surveys ([3], [4] and [29]). 
This program allowed us to find some peculiar sources, e.g. Swift J0739.6$-$3144 that is a possible Compton thick source [25] or PBC J2333.9$-$2343, a source with 3 different classifications in 3 different wavebands (radio, optical and X-ray; [26]). Moreover, the X-ray observations enabled us to correlate X-ray versus optical classification and discuss peculiar sources [26].

In the latest {\it Swift}/BAT paper ([27]) we reduced and analyzed the optical spectra of 76 previously unidentified BAT sources, obtaining the redshift, distance and emission line fluxes. Preliminary results show that 38 objects are
Seyfert 1, 24 Seyfert 2, 2 XBONGs, 5 Pure LINERs, 1 Seyfert 2/LINERs and 6 CVs. 
For the sake of completeness in Fig, \ref{histo}, right panel, we reported the redshift distribution of all the extragalactic objects (70 sources) belonging to this new {\it Swift} sample. All the objects are placed in the nearby Universe (z $<$ 0.3), except one at redshift 1.137.

\section{Conclusions}
Our group has either given for the first time, or confirmed, or corrected, the optical spectroscopic identification of more than 300 sources from the {\it INTEGRAL} and {\it Swift} surveys. This was achieved through a multisite observational campaign in Europe, Africa, Central and South America. The majority of these sources are AGNs, followed by X-ray Binaries (XRBs). This trend is reflected also in the latest identification papers presented in this work; for the Swift identifications we found 6 CVs and 70 AGNs, all placed in the local Universe at z $<$ 0.3 except one at 1.137, while for the INTEGRAL identifications we have 6 XRBs and 17 AGNs, 6 of them placed at z $>$ 0.3. 
As a final remark, all of the results shown in this work stress the importance of the optical spectroscopic followup for the classification of unidentified sources and for the study of their statistical properties.

\end{document}